\documentclass[aps,groupedaddress,nofootinbib]{revtex4}
\usepackage{amsmath}
\usepackage{graphicx}
\usepackage{amsfonts}
\usepackage{amssymb}%
\providecommand{\U}[1]{\protect\rule{.1in}{.1in}}

\newcommand{\BOX}{\hbox {$\sqcap$ \kern -1em $\sqcup$}}

\newcommand{\be}{\begin{equation}}
\newcommand{\ee}{\end{equation}}
\newcommand{\ba}{\begin{eqnarray}}
\newcommand{\ea}{\end{eqnarray}}
\newcommand{\ban}{\begin{eqnarray*}}
\newcommand{\bea}{\begin{eqnarray}}
\newcommand{\eea}{\end{eqnarray}}
\newcommand{\ean}{\end{eqnarray*}}
\newcommand{\barr}{\begin{array}}
\newcommand{\earr}{\end{array}}

\begin{document}
\title{Cosmological Constant and Local Gravity }
\author{Jos\'e Bernab\'eu~}
\affiliation{Departamento de F\'isica Te\'orica and IFIC, Universidad de
Valencia-CSIC, E-46100, Burjassot (Valencia), Spain,
and \\ CERN, Theory Division, CH-1211 Geneva 23, Switzerland.}

\author{Catalina Espinoza}
\affiliation{Departamento de F\'isica Te\'orica and IFIC, Universidad de
Valencia-CSIC, E-46100, Burjassot (Valencia), Spain.}

\author{Nick E. Mavromatos}
\affiliation{King's College London, University of London, Department of Physics, Strand
WC2R 2LS, London, U.K.}

\begin{abstract}
{\small  We discuss the linearization of Einstein equations in the presence of a cosmological constant, by expanding the solution for the metric around a flat Minkowski space-time. We demonstrate that one can find consistent solutions to the linearized set of equations for the metric perturbations, in the Lorentz gauge, which are not spherically symmetric, but they rather exhibit a cylindrical symmetry. We find that the components of the gravitational field  satisfying the appropriate Poisson equations have the property of ensuring that a scalar potential can be constructed, in which both contributions, from ordinary matter and $\Lambda > 0$, are attractive. In addition, there is a novel tensor potential, induced by the pressure density, in which the effect of the cosmological constant is repulsive. We
also linearize the Schwarzschild-de Sitter exact solution of Einstein's equations (due to a generalization of Birkhoff's theorem) in the domain between the two horizons. We manage
to transform it first to a gauge in which the 3-space metric is conformally
flat and, then, make an additional coordinate transformation leading
to the Lorentz gauge conditions. We compare our non-spherically symmetric solution
with the linearized Schwarzschild-de Sitter
metric, when the latter is transformed to the Lorentz
gauge, and we find agreement. The resulting metric, however, does not acquire a proper Newtonian form in terms of the unique scalar potential that solves the corresponding Poisson equation. Nevertheless, our solution is stable, in the sense that the physical energy density is positive. }

\end{abstract}
\maketitle


\section{INTRODUCTION}

   De Sitter space has become physically relevant in the last decade,
after the extraordinary discovery, based on astrophysical and
cosmological observations, that our Universe is accelerating at late
eras. A diverse range of results~\cite{1}, from cosmic microwave background
temperature fluctuation measurements to high-redshift supernova
measurements, baryon acoustic oscillation measurements and weak lensing techniques
(such as cosmic shear and red-shift space distortion), has
indicated that a global best fit for Cosmology is provided by a simple
Friedman-Robertson-Walker Universe with a positive cosmological
constant $\Lambda$~\cite{2}, whose dominance over matter at late eras is held
responsible for the Universe acceleration. In fact, according to this
simple and successful scenario, the best fitting of the data says that
over 70 \% of the current-epoch Universe energy-density budget is
dominated by this mysterious form of dark energy, which is compatible
with a cosmological constant $\Lambda$.

   Once $\Lambda$ is included in the Einstein equations for General Relativity,
its implications are not limited to cosmological problems. The
observable effects of the cosmological constant for orbits in the Solar
system and double pulsars exist~\cite{3}, but are too small to be detected. Its effect becomes
more important for extended galaxy clusters through a correction to
the Virial relation \cite{4}. In this paper we address the problem of
understanding the consistency of the theory, in the presence of $\Lambda$,
when applied to local gravity and whether the steps of linearizing
gravity and going to the Lorentz gauge lead to a sound description of
the gravitational field within special relativity, and to a Newtonian
form for the metric, in terms of a unique scalar potential. The
prevalent view ~\cite{5} is that the effects of the cosmological constant
are equivalent locally, \emph{i.e.}, within the distances of galaxies or
galaxy clusters, to those corresponding to a repulsive tidal force,
of a conservative nature, being derived from a unique scalar potential
$\Phi_\Lambda$ of the form
\begin{equation}\label{eq1}
                      \Phi_{\Lambda}/c^2  =  - \frac{1}{6}\Lambda r^2
\end{equation}
The Correspondence Principle demands that the theory should contain
the Newtonian limit for weak classical gravity, with a relative motion
of the source much smaller than \emph{c} and with the material stresses much
smaller than the mass-energy density. In absence of $\Lambda$, General
Relativity satisfies this limit. The requirement of the Newtonian limit
is made explicit by linearizing the theory for weak gravity, expanding
around the flat Minkowski space-time $\eta_{\mu\nu} = \left (1,\, -1,\,-1,\,-1\right)$, writing
in Cartesian components
\begin{equation}\label{eq2}
                      g_{\mu\nu} = \eta_{\mu\nu} + h_{\mu\nu}
\end{equation}
and assuming that both $h_{\mu\nu}$ and the derivatives are small with respect to unity, so only first
order terms are kept. With the transformation to the Lorentz gauge and the dominance
of the mass-energy density for the gravity source, one derives the
"\emph{Newtonian metric}" for the gravitational field
\be\label{eq3}
   ds^2 = (1 + 2 \Phi/c^2) c^2\,dt^2 - (1 - 2\Phi/c^2)\, d\vec{x}^2
\ee
where $\Phi$ is the unique scalar potential, obtained as a solution of the
Poisson equation with the appropriately linearized source. One should
notice that the 3-space metric in (\ref{eq3}) is \emph{conformally flat}. If these
conditions are met, the constant $\kappa$ of General Relativity in the
Einstein equations, $ R_{\mu\nu} - \frac{1}{2}g_{\mu\nu} R = -\kappa T_{\mu\nu}$,
describing the coupling of the source to the
gravitational field, is identified with Newton constant G as
\be\label{eq4}
                \kappa(\Lambda=0) = 8 \pi G / c^4
\ee
In using (\ref{eq3}) and (\ref{eq4}) for the geodesics of a test particle in
non-relativistic motion, the result is Newton equation of motion
with a force  $\vec{F} = - \vec{\nabla} \Phi$. In that non-relativistic
limit, only the $h_{00}$ component of the gravitational field is operative.
Eq. (\ref{eq3}), however, is more general and it applies, for example, to the
problem of light bending within Special Relativity.

     How is this picture modified, if any, by the presence of the (positive)
cosmological constant $\Lambda \ne 0$ ? It is quite frequent to see in the
literature that the above results are preserved and, then, one adds the
$\Lambda$-term \emph{a posteriori}. We emphasize, however, that the conditions
to arrive to (\ref{eq3}) are \emph{not guaranteed} in the presence of $\Lambda$ and, in
particular, that the theory contains two fundamental constants: $\kappa$
and $\Lambda$. Without the Newtonian limit, the identification (\ref{eq4}) finds no
justification.

     The language of a gravitational field necessitates the introduction of a space-time background
metric.
For the purpose of this work, the background will
still be Minkowski's, so that weak gravity means first order in both
$\kappa$ and $\Lambda$. In this view, $\Lambda$ is sitting in the right-hand side of
Einstein equations, with the meaning of an unavoidable gravity source
as "dark energy". We first find in Section \ref{sec2} the new solutions for
Linearized Gravity assuming that the conditions for the choice of
coordinates in the Lorentz gauge are satisfied. We shall see that
we have to depart from strict spherical symmetry, finding some
components of the field with cylindrical symmetry. In Section \ref{sec:3} we
linearize the Schwarzschild-de Sitter solution and proceed, first,
to transform it to a gauge in which the 3-space metric is conformally
flat and, then, make an additional coordinate transformation leading
to the Lorentz gauge conditions. We compare the solution
found in Section \ref{sec2} with that for the linearized Schwarzschild-de Sitter
metric~\cite{6} when the latter is appropriately transformed to the Lorentz
gauge. We investigate whether the resulting metric is Newtonian, i.e.,
with the structure shown in (\ref{eq3}) in terms of a unique scalar potential.
In the physically meaningful coordinate frame where the metric of 3-space is conformally flat, we examine effects to first order in both $\Lambda$  and $m$, where $m$ is the mass of a massive celestial
object, on the photon orbit, thus demonstrating the existence of non-trivial effects of the cosmological constant on it.
These findings are relevant for the on-going debate in the literature about the r\^ole of $\Lambda$ on the deflection
of light. Finally, Section \ref{sec4} presents a discussion of the results and some open issues.

\section{LINEARIZED GRAVITY IN THE PRESENCE OF A COSMOLOGICAL CONSTANT\label{sec2}}

    The Newtonian gravitational field is described by a single scalar
potencial $\Phi(\vec{x})$ which is a solution of the Poisson equation,
with a matter source given by the mass density $\rho_m$
\be\label{eq5}
                 \nabla^2 \Phi(\vec{x})  =   4 \pi G \rho_m
\ee
    The coupling of the source to the gravitational field is determined
by the Newton constant $G$.

    Einstein equations, derived from the Einstein-Hilbert
four-dimensional action, read
\be\label{eq6}
        R_{\mu\nu} - \frac{1}{2}g_{\mu\nu}R + \Lambda g_{\mu\nu} = - \kappa T_{\mu\nu}
\ee
where $\Lambda > 0$  is the cosmological constant, which has the dimensions of a
curvature, namely [length]$^{-2}$, $T_{\mu\nu}$ is the stress-energy tensor
of matter in the gravitational field generated by the metric tensor
$g_{\mu\nu}$ and $\kappa$ is a dimensionful constant defining the coupling of
gravity to matter.

    Clearly Eqs.(\ref{eq6}) do \emph{not} permit a flat spacetime in the absence of
matter sources, when they reduce to
\be\label{eq7}
                    R_{\mu\nu} = \Lambda g_{\mu\nu}
\ee
which implies curvature. In spite of this, we still insist on the
possibility of connecting (\ref{eq6}) with a linearized theory around Minkowski
flat space-time, within Special Relativity, using the expansion (\ref{eq2}):
$h_{\mu\nu}$ would be the metric perturbation, a tensor under global
Lorentz coordinate transformations, This approach necessarily implies
that the cosmological constant $\Lambda$ is \emph{not} a mere geometric effect but has to be put on the
right-hand side of (\ref{eq6}) and interpreted as a form of \emph{dark energy}, \emph{i.e.},
as an additional source of gravity beyond matter. The consistency of the
linearized theory requires the limit of \emph{weak gravity},
with the source being of first order, as is the field $h_{\mu\nu}$ and its
derivatives. As a consequence, we have to stay in first order in both
$\kappa$ and $\Lambda$, without any back-reaction of the gravitational field on
$T_{\mu\nu}$ and $g_{\mu\nu}$ in the right-hand side of Eq.~(\ref{eq6}).

    Upon defining the trace-reversed version of the $h$'s as
\be\label{eq8}
               \tilde{h}_{\mu\nu} = h_{\mu\nu} - \frac{1}{2}\eta_{\mu\nu} h~,
\ee
it follows that:
\bea\label{eq9}
      &~& h \equiv h^\alpha_{\,\,\alpha}  =-\tilde{h}^\mu_{\,\,\mu} \equiv-\tilde{h} \nonumber  \\                                                              &~& h_{\mu\nu} = \tilde{h}_{\mu\nu}- \frac{1}{2}\eta_{\mu\nu} \tilde{h}~,
\eea
where it is understood the indices are raised and lowered with the Minkowski metric.
One observes that the Ricci tensor $R_{\mu\nu}$ can be written as
\be\label{eq10}
        R_{\mu\nu} = \frac{1}{2}\left(\Box h_{\mu\nu} - h_{\, ,\mu\nu} - h^\lambda_{\,\,\mu,\nu\lambda} - h^\lambda_{\,\,\nu,\mu\lambda}\right)~,
\ee
to leading order in the field and its derivatives, with $\Box \equiv \eta^{\lambda\rho}\partial_\lambda\partial_\rho$
denoting the D'Alembertian operator.

    A suitable coordinate transformation can get rid of the last two
terms of Eq. (\ref{eq10}) if we choose the following condition
\be\label{eq11}
                 \tilde{h}^\mu_{\,\,\nu,\,\mu} = 0~.
\ee
This is called the \emph{Lorentz gauge}, in analogy with electromagnetism.

   In the Lorentz gauge, the Einstein tensor
$G_{\mu\nu}=R_{\mu\nu}-\frac{1}{2}g_{\mu\nu}R$ becomes
\be\label{eq12}
                  G_{\mu\nu} = \frac{1}{2}\Box \tilde{h}_{\mu\nu}~,
\ee
so that the full field equations (\ref{eq6}) now reduce to
\be\label{eq13}
         \Box \tilde{h}_{\mu\nu} = -2 \kappa T_{\mu\nu}- 2\Lambda \eta_{\mu\nu}
\ee
Eqs.(\ref{eq13}), plus the gauge condition (\ref{eq11}), constitute the basic
ingredients for our study. These linear field equations are also
decoupled when they are written in terms of $\tilde{h}_{\mu\nu}$. As a consequence, the trace-reversed
$\tilde{h}_{\mu\nu}$ has the meaning of the gravitational field and Eq.~(\ref{eq13}) is the relativistic generalization of the Newtonian gravity equation (\ref{eq5}).

        We are interested in the solution for the linearized metric,
in the Lorentz gauge, for a weak static field generated by a point mass
$M$ located at $r = 0$: all components of $T_{\mu\nu}$ vanish except
\be\label{eq14}
                    T_{00} = M c^2 \delta(\vec{x})
\ee
In terms of   $m \equiv \kappa M c^2 / 8 \pi$, we find the static solutions to
(\ref{eq13}) and (\ref{eq11}) in  diagonal form in the presence of $\Lambda$. Each diagonal
component $\tilde{h}_{\mu\mu}$ (no sum over $\mu$) is independent of the specific
coordinate $x^\mu$. This condition implies that the space
components of the gravitational field  $\tilde{h}_{i i}$ (no sum over $i$),
if they exist, have to depart from spherical symmetry. The solution
is
\be\label{eq15}
\tilde{h}_{\mu\nu} =
\left(\begin{matrix}
  -\frac{4m}{r} + \frac{1}{3}\Lambda r^2 &  0 & 0  & 0 \\
                          0    &   - \frac{1}{2}\Lambda (y^2 + z^2) & 0 &  0 \\
                           0 &  0  &                - \frac{1}{2}\Lambda (z^2 + x^2) &  0 \\
 0 & 0 & 0 & - \frac{1}{2}\Lambda (x^2 + y^2)  \end{matrix}\right)
\ee
We observe that these components of the field, each of them solution
of Poisson's equations, satisfy the following:
\begin{itemize}
\item{1)} The scalar potential
$\Phi = \frac{1}{4}c^2 \tilde{h}_{00}$  gets modified by an \emph{attractive} \emph{effect} induced
by a \emph{positive energy density} due to $\Lambda$, besides that generated by (\ref{eq14}) due to ordinary matter;

\item{2)} there is a \emph{novel tensor potential}
$\tau_{ij} = \frac{1}{4}c^2 \tilde h_{ij}$ with a \emph{repulsive effect} induced by the
\emph{negative} \emph{pressure density} due to $\Lambda$;

\item{3)} whereas the scalar potential presents
a rotational symmetry, the components of the tensor potential have
a \emph{cylindrical symmetry} around the corresponding principal axis. The
cylindrical symmetry is, in fact, the same for the three principal
axes, a remnant of the rotational symmetry. This \emph{breaking} of the
rotational symmetry is an \emph{artifact} of the gauge fixing imposed by the
Lorentz condition.
\end{itemize}
      From Eq.~(\ref{eq15}) we find the complete trace of the field and the
3-space trace of the tensor potential as
\begin{eqnarray}\label{eq16}
    &&           \tilde{h}^\alpha_{\,\, \alpha} \equiv \tilde{h} = - \frac{4m}{r} + \frac{4}{3} \Lambda r^2 \nonumber \\
&&             \tau^i_{\,\,i} \equiv \tau = + \frac{1}{4}c^2 \Lambda r^2~.
\end{eqnarray}

From these results it becomes already apparent that the metric  is \emph{not Newtonian}
when $\Lambda \ne 0$. In going from the trace-reversed gravitational field $\tilde{h}_{\mu\nu}$
to the metric perturbation in the Lorentz gauge, we find
\be\label{eq17}
 h_{\mu\nu} = \left(\begin{matrix}
2 \Phi - 2 \tau  & 0 & 0 & 0   \\
0 & 2 \Phi + 2 \tau + 4 \tau_{11} & 0 &  0 \\
 0 & 0  & 2\Phi + 2 \tau + 4 \tau_{22} & 0 \\
0 & 0 & 0 & 2 \Phi + 2 \tau + 4 \tau_{33}\end{matrix}\right)
\ee
with the property  $h = - \tilde{h}$ for the complete trace, as required.
Eq. (\ref{eq17}) gives the solution for the metric within special relativity.
It is neither conformally flat nor of the form we have  for $\Lambda$ = 0, i.e.  $g_{00} = 1 + 2 \Phi/c^2$,
$g_{ii} = -1 + 2 \Phi/c^2$ (no sum over $i$), $i=1,2,3$. The contribution of the modified
scalar potential $\Phi$ to the metric (\ref{eq17}) is still as in the
Newtonian form (\ref{eq3}), but all the diagonal components get additional contributions
from the tensor potential. We realize, in particular, that the $h_{00}$
component acquires \emph{an effective} $\Lambda$-\emph{term}, which originates from both
the attractive scalar potential and the trace of the repulsive tensor
potential: its net effect is repulsive and given by (\ref{eq1}). One should
notice that, whereas
\be\label{eq18}
                        h_{00} = - 2\frac{m}{r} - \frac{1}{3}\Lambda r^2~,
\ee
is the only component of the metric entering the geodesic equation for
the motion of a non-relativistic body, given  by $\frac{d^2 x^i}{d t^2} \simeq -c^2\Gamma^i_{00}$, where $\Gamma_{\mu\nu}^\rho$ is the Christoffel symbol, the 3-space components of
the metric
\be\label{eq19}
                     h_{ii} = - 2\frac{m}{r} + \frac{1}{6}\Lambda (r^2 + 3 x_i^2)~, \quad i=1,2,3 \quad (\rm{no~sum~over~i})
\ee
are also intervening for a test-body in relativistic motion. For the
case of light bending, we shall discuss the effects of the cosmological constant $\Lambda > 0$ on the photon orbit in section~\ref{sec:3}. We note at this point that within the Schwarzschild-de Sitter metric, the magnitude of the effects of the cosmological constant $\Lambda > 0$ on the bending of light from distant galaxies
is still an open issue. At present, there are several approaches to the subject, giving different answers, in which the order of magnitude of the effects ranges from zero~\cite{8}  or unobservably small~\cite{10}
to appreciable one~\cite{7,9}.

Our considerations in this paper pertain to local effects of the cosmological constant, which are discussed within the Special Relativity framework. As we show in the next section \ref{sec:3}, our linearized solution (\ref{eq17}) can be obtained from the Schwarzschild-de Sitter solution upon appropriate coordinate transformations.

\section{THE LINEARIZED SCHWARZSCHILD-DE SITTER SOLUTION\label{sec:3}}

 At this point one should investigate the connection of the linearized gravity solution found in
Section \ref{sec2} with the Schwarzschild-de Sitter metric. We do know that,
under the conditions we have imposed for the source of coupling with
gravity, there is  a theorem analogous to \emph{Birkhoff's} for the
Schwarzschild metric, stating~\cite{6} that there is a unique static
solution with spherical symmetry of the form~\footnote{According to the analysis in \cite{6}, if one relaxes the assumption of staticity, the spherical symmetry implies two types of solutions in the case of positive $\Lambda > 0$:
{\bf (i)} a \emph{static} solution (\ref{eq20}), expressing the field around a spherically symmetric mass,
where, as in the Schwarzschild case, the mass is an integration constant of the Einstein equations;
{\bf (ii)} a \emph{non-static}, time-dependent solution, distinct from that due to the field around a spherical distribution of masses, which consists of successive identical spheres, that is, a \emph{cylindrical} Universe of Bertotti-Kasner type:
    \begin{equation}
  ds^2 = dt^2 - e^{2\sqrt{\Lambda} \, t}\,dr^2 - \frac{1}{\Lambda} \left(d\theta ^2 + {\rm sin}^2\theta d\phi^2 \right)~,
\label{bksol}
\end{equation}
In view of the last term, the reader should understand now why this solution exists only in the case of positive $\Lambda$.}:
\be\label{eq20}
 ds^2= \left(1-2\frac{m}{\bar{r}}-\Lambda\bar{r}^2/3\right) c^2 dt^2 - \left(1-2\frac{m}{\bar{r}}-\Lambda\bar{r}^2/3\right)^{-1} d\bar{r}^2
          -\bar{r}^2 (d\bar{\theta}^2 +  \rm{sin}^2\bar{\theta} d\bar{\phi}^2)
\ee
This is the Schwarzschild-de Sitter metric, in which the Schwarzschild
space coordinates define $\bar{r}$ as the "\emph{area distance}", \emph{i.e.}, the
distance for which the surface is given by the Euclidean measure
$4 \pi \bar{r}^2$. The presence of horizons at the approximate values
$\bar{r} = 2m$ and $\bar{r}=\sqrt{3/\Lambda}$ should be noted along with the fact that the observer
is actually required to live in the space between them, for our \emph{weak-gravity} analysis to be valid.
     Indeed, in the domain
\be\label{eq21}
                2m \ll  \bar{r}  \ll  \sqrt{3/\Lambda}
\ee
we can linearize the components of the metric (\ref{eq20}) around Minkowski background space-time.
Even for the case $\Lambda =0$ and a fortiori for $\Lambda \ne 0$,
the result for the  metric (\ref{eq20}) is neither conformally flat nor of the
Newtonian form (\ref{eq3}) with a single scalar potential. We notice, in particular,
that the lack of a conformally flat metric in these Schwarzschild
coordinates implies that the metric is not diagonal when written in
Cartesian components, even if it is so in spherical components. One can
check that the corresponding linearized metric does \emph{not} satisfy the
\emph{Lorentz gauge} condition.

      At this point it is natural to ask whether it is possible to reach the Lorentz gauge by a coordinate
transformation. To this end, we proceed in two steps:

(i) First, we move from the Schwarzschild coordinates to fully symmetric
spherical coordinates leading to a conformally flat metric in 3-space.
This transformation exists because, with $\Lambda \ne 0$, we are in a case of \emph{constant
curvature}. With the transformation
\be\label{eq22}
              \bar{r}  \rightarrow  r' = \bar{r} (1 - \frac{m}{\bar{r}} + \frac{1}{12}\Lambda \bar{r}^2)
\ee
the resulting metric becomes
\be\label{eq23}
ds^2 = (1 - 2\frac{m}{r'} - \frac{1}{3}\Lambda {r'}^2) c^2 dt^2 - (1 + 2\frac{m}{r'} - \frac{1}{6}\Lambda {r'}^2)d{\vec{x'}}^2
\ee
In Eq. (\ref{eq23}), both cartesian and spherical components of the 3-space
metric are diagonal. In spite of this property, the metric is not of the
form of Eq. (\ref{eq3}) in terms of the $r'$ coordinate, \emph{i.e.} $(1 + 2\Phi'(r'))$ vs. $(1 - 2\Phi'(r'))$,  if $\Lambda \ne 0$. One can check, in
fact, that the \emph{Lorentz gauge} condition is \emph{not} satisfied for the metric (\ref{eq23})
in the presence of a non-zero cosmological constant, $\Lambda \ne 0$,  which is to be contrasted with  the $\Lambda=0$ case, for which
the two gauges coincide.
The trace of the linearized metric perturbation is given by
\be\label{eq24}
                     h' = 4\frac{m}{r'} - \frac{5}{6} \Lambda {r'}^2
\ee
so that the trace-reversed gravitational field (\ref{eq8}) is given, in these
coordinates, by
\be\label{eq25}
\tilde{h}'_{\mu\nu} =\left(\begin{matrix}
                  - 4\frac{m}{r'} + \frac{1}{12}\Lambda {r'}^2 & 0 & 0 & 0 \\
                         0 & -\frac{1}{4} \Lambda {r'}^2 & 0 & 0 \\
   0 & 0 & -\frac{1}{4} \Lambda {r'}^2 & 0 \\
  0 & 0 & 0  & - \frac{1}{4}\Lambda  {r'}^2 \end{matrix} \right)
\ee
Notice that, for $\Lambda = 0$, this field would be described by a unique
scalar potential solution of the Laplace equation for $r' > 0$.
However, for $\Lambda \ne 0$, the field components do not satisfy the Lorentz
gauge condition for the coordinates $(x', y', z')$.

In this fully symmetric gauge, the scalar, $\Phi'$, and tensor,
$\tau'_{ij} = \tau' \eta_{ij}/3$, potentials, with $\tau'$ the corresponding 3-space trace,
are given by
\begin{eqnarray}\label{eqn26b}
&&               \Phi'/c^2 = - \frac{m}{r'} + \frac{1}{48}\Lambda {r'}^2 \nonumber \\
&&             \tau'/c^2 = + \frac{3}{16} \Lambda {r'}^2~.
\end{eqnarray}
They satisfy the following coupled Equations in terms of the sources
\begin{eqnarray}\label{eq27b}
&&       {\nabla'}^2 (\Phi' + \frac{1}{3}\tau') = \frac{1}{2} \kappa c^2 T_{00} + \frac{1}{2}\Lambda c^2~, \nonumber \\
&&           {\nabla'}^2 \tau' = \frac{9}{8} \Lambda c^2~.
\end{eqnarray}
The corresponding 3-space conformally flat metric (\ref{eq23}) can then be
written in the simple form
\be\label{eq28b}
 ds^2
= (1 + \frac{2(\Phi'-\tau')}{c^2})c^2 dt^2- (1- \frac{2\Phi'}{c^2}- \frac{2\tau'}{ 3c^2})d\vec{x'}^2~.
\ee
This coordinate frame is the appropriate one for measurements using standard clocks and rods, due to the fully isotropic
3-space metric components. In the case of photons, the null geodesics, derived from (\ref{eq23}), can be easily constructed, leading to a direction-independent speed of light, a feature that is not valid on other coordinate frames..
 The geodesics encompass, as usual, the two first integrals of motion related to the photon energy and angular momentum. The orbit ${r'}(\phi, \theta=\frac{\pi}{2})$ can be expressed as an equation for the
function $u (\phi) \equiv \frac{R}{r'}$, which to first order in $m$ and $\Lambda$ reads:
\begin{equation}
\left(\frac{du}{d\phi}\right)^2 + u^2 -1 = \frac{4m}{R} \left(u -1\right) + \frac{1}{6}\Lambda R^2 \left(\frac{1}{u^2}-1\right)~.
\label{orbitu}
\end{equation}
where $R$ is the radius of the spherical mass distribution and we consider a path of the photon that grazes its surface
at $\phi=\frac{\pi}{2}$. A symmetric solution about the axis $\phi=\pi/2$, as required by the geometry, can be found analytically around $\phi \simeq 0$ and $\phi \simeq \pi$, to leading order in ${\rm sin}\phi \ll 1$ for the first-order terms proportional to $m$ and $\Lambda$:
\begin{equation}\label{orbit2}
 u(\phi) \simeq {\rm sin}\phi + \frac{2m}{R} - \frac{1}{12} \frac{\Lambda R^2}{{\rm sin}\phi}~.
\end{equation}
For the case $\Lambda = 0$, there is an asymptotic $u \to 0$ solution, leading to the one-sided bending angle $\phi_\infty = -\frac{2m}{R}$, thus reproducing
the standard Einstein's result in General Relativity.

However, in the presence of the $\Lambda$-term, there is no asymptotic limit, due to the de-Sitter horizon.
As we observe from Eq.~(\ref{orbit2}), there are non-trivial effects of order $\Lambda$ to the photon orbit in our 3-space-conformally-flat coordinate system, which come with opposite sign to the $\Lambda =0$ General Relativity contribution. Indeed,
the one-sided bending angle $\beta_\Lambda (r')$ obtained from (\ref{orbit2}), with $\Lambda > 0$, is:
\begin{equation}\label{bendangl}
   \beta_\Lambda (r') \simeq -\frac{2 m}{R} + \frac{1}{12}\Lambda R {r'}~, \qquad {r'} \gg R~.
\end{equation}
Given that a conformal transformation of the 3-space coordinates preserves the angles, the above result would also be valid in flat space-time, which is the background space-time in our linearized gravity approximation.
On the other hand, as shown in ref.~\cite{islam}, if one writes the
orbit of the photon in the original (non-conformal) Schwarzschild coordinates, the orbit appears to be independent of $\Lambda$.

Our conclusion that $\Lambda$ is contributing through Eq.~(\ref{bendangl}) to the deflection of light by a spherical mass distribution is in line with the claims~\cite{10,7,9} that the effect exists, in spite of the $\Lambda$-independent equation in Schwarzschild coordinates. A more detailed analysis of this and other observational effects of a gravitating $\Lambda$ will be presented elsewhere.

(ii) As a second step, we make the coordinate transformation
\bea\label{eq26}
    &&           x' \rightarrow  x = x' + \frac{1}{12}\Lambda {x'}^3 \nonumber \\
&&               y' \rightarrow y = y' + \frac{1}{12}\Lambda {y'}^3 \nonumber \\
&&             z' \rightarrow z = z' + \frac{1}{12}\Lambda {z'}^3~,
\eea
which definitively leads to the Cartesian components of the field (\ref{eq15})
in the coordinates $(ct, x, y, z)$. These ones are the coordinates
associated with the Lorentz gauge and the discussion after Eq. (\ref{eq15})
follows. We conclude that the solution found in Section \ref{sec2} is precisely
the linearized Schwarzschild-de Sitter metric, written in a new set of
appropriate coordinates that correspond to the Lorentz gauge.

\section{Conclusions\label{sec4}}

In this work we have discussed a first-order solution to the Einstein equations in the presence of a small positive
cosmological constant $\Lambda > 0$, in the case where the equations are linearized about flat Minkowski space-time.
Usually, in the literature, this problem is associated straightforwardly with the presence of a repulsive force in the ``Newtonian limit''. As we emphasized in this note, however, since there is no proper Newtonian metric in the
case $\Lambda \ne 0$, the linearization procedure has to be applied with care, and the identification of what plays the r\^ole of a repulsive ``potential'' presents subtleties. In particular,
we have found that in this case there are scalar and tensor potentials, which \emph{both} describe
the gravitational field in special relativity in the presence of a non-trivial (no matter how small) Cosmological Constant. As we have discussed,
with $\Lambda \ne 0$ there is \emph{no Newtonian form} for the metric in any gauge.
The Lorentz gauge, for $\Lambda \ne 0$, does not lead to a 3-space conformally flat metric.

If one insists
on the requirement of a Newtonian form for the metric, in agreement with the Correspondence Principle,
then one has to conclude that $\Lambda$ cannot be a \emph{classical geometric} \emph{effect}, but rather a \emph{relativistic quantum effect}, which should vanish (formally) for $\hbar \rightarrow  0$. From dimensional arguments,
by writing  $\Lambda  = \kappa \rho_\Lambda$, the dark energy density $\rho_\Lambda$ is given by
\be\label{eqfinal}
                         \rho_\Lambda \sim   \hbar c / \ell^4
\ee
where $\ell$ should be a characteristic length, whose microscopic origin, and hence its order of magnitude, is  a mystery. In this case, the relation (\ref{eq4}) of $\kappa$ with Newton constant $G$ would still be valid for $\Lambda \ne 0$.

Our discussion in this note has been concentrated solely on local effects of the cosmological constant.
At this point we feel that we should contrast our findings with the traditional point of view adopted in the literature, in which a ``Newtonian scalar potential'' is constructed from the $h_{00}$ component of the metric perturbation by looking only at the Non-relativistic geodesics, \emph{i.e.} from Eq.~(\ref{eq18}). Using the metric
for the Schwarzschild-de Sitter solution (\ref{eq20})~\cite{5,6},  in the limit of $m \to 0$, or equivalently in the case where the $\Lambda$ term dominates over the
mass term, it is often stated that the resulting ``energy density'' appears \emph{negative}, thereby indicating an instability. Indeed, this line of thinking prompted the elevation of the Schwarzschild-de Sitter solution from a local to a global one, relevant for an expanding Universe, where this instability is remedied, in the sense that the cosmological de Sitter solution has still positive energy, but negative pressure.

However, as our analysis in this article has shown (c.f. discussion following eq.~(\ref{eq15})), this ``instability'' arises from a misinterpretation of what energy density means in the linearized solution. In our case, precisely due to the absence of a proper Newtonian form of the metric, the correct identification of the energy density can only be made via the Poisson equation for the gravitational field $\tilde{h}_{\mu\nu}$, eq.~(\ref{eq13}). From this point of view,  our solution appears \emph{stable}. The corresponding dark energy density is indeed proportional to $\Lambda > 0$, and originates from an attractive contribution to the scalar potential defined via $\tilde{h}_{00}$ in (\ref{eq15}), even in the case $m \ll \Lambda$ (\emph{i.e.} when $|\kappa T_{\mu\nu}| \ll |\Lambda \eta_{\mu\nu}|$ in (\ref{eq13})), and hence is positive.

A final comment before closure. The cosmological de Sitter solution, in the context of a Friedman-Robertson-Walker (FRW) Universe, is an entirely different problem. The precise connection from the global solution to the local one, by means of space-dependent cosmological perturbations, is a complicated issue, which is still unsolved. One might hope that, by perturbing the $\Lambda$-de Sitter-FRW appropriately, it would be possible to make a connection with the local gravity case and thus compare the results with the analysis presented here.

\section*{Acknowledgements}

We wish to thank D. Espriu, V.A. Mitsou, M. Quir\'os, R. Rebolo and M. Sereno for interesting discussions.
We also acknowledge helpful suggestions and comments by the anonymous referee.
The research work of J.B. and C.E. is supported in part by the Grants of the Spanish Ministry of Science and Innovation FPA 2008-02878 and of the Valencia Generalitat PROMETEO 2008/004, while that of N.E.M. is partly supported by the European Union, through the FP6 Marie
Curie Research and Training Network \emph{UniverseNet} (MRTN-CT-2006-035863).

\end{document}